**Intrinsic low-spin state and strain-tunable anomalous Hall scaling in high-quality SrRuO$_3$ (111) films**

Harunori Shiratani,[1,2] Yuki K. Wakabayashi,[1,a] Yoshiharu Krockenberger,[1] Masaki Kobayashi,[1] Kohei Yamagami,[3] Takahito Takeda,[4] Shinobu Ohya,[2,5] Masaaki Tanaka,[2,5] and Yoshitaka Taniyasu[1]

[1]Basic Research Laboratories, NTT, Inc., Atsugi, Kanagawa 243-0198, Japan

[2]Department of Electrical Engineering and Information Systems, The University of Tokyo, 7-3-1 Hongo, Bunkyo, Tokyo 113-8656, Japan

[3]Japan Synchrotron Radiation Research Institute (JASRI), 1-1-1 Kouto, Sayo, Hyogo, 679-5198, Japan

[4]Department of Chemical System Engineering, The University of Tokyo, 7-3-1 Hongo, Bunkyo, Tokyo 113-8656, Japan

[5]Center for Spintronics Research Network, The University of Tokyo, 7-3-1 Hongo, Bunkyo, Tokyo 113-8656, Japan

a)Author to whom correspondence should be addressed: yuuki.wakabayashi@ntt.com

**ABSTRACT**

The (111)-oriented 4$d$ ferromagnetic perovskite SrRuO$_3$ (SRO) offers a unique triangular-lattice geometry, making it a promising platform for exploring Berry-curvature-driven and spin-orbit-coupled transport. Here, we present a systematic study of the structure, magnetism, and magnetotransport of high-quality SRO (111) thin films with thicknesses $t$ = 1.2–60 nm grown on SrTiO$_3$ (111) substrates by machine-learning-assisted molecular beam epitaxy. We achieved a residual resistivity ratio of 45.5 in a 60 nm-thick film, the highest reported for this orientation, enabling access to intrinsic electronic and magnetic behavior. Temperature-dependent resistivity confirms Fermi-liquid transport below ~15 K in both coherently strained ($t$ = 10, 20 nm) and strain-relaxed ($t$ = 60 nm) films, thereby enabling detailed magnetotransport and magnetic measurements. The linear, non-saturating positive magnetoresistance persists up to 14 T, while Hall-effect measurements and temperature scaling separate intrinsic (Karplus–Luttinger) and extrinsic (side-jump) contributions to the anomalous Hall effect, with the relative weight tuned by (111) epitaxial strain. X-ray magnetic circular dichroism at the Ru $M_{2,3}$ and O $K$ edges, together with SQUID magnetometry, demonstrates an intrinsically low-spin Ru ground state for both coherently strained and relaxed films, resolving ambiguities among prior reports. These detailed crystalline, electrical, and magnetic characterizations provide a rigorous foundation for understanding and engineering quantum transport in SRO (111).

## I. INTRODUCTION

SrRuO$_3$ (SRO) is a 4$d$ transition-metal perovskite oxide that has been extensively studied as a prototypical itinerant ferromagnet with intermediate electron correlation[1–3] and large magnetocrystalline anisotropy.[4,5] Owing to its high metallic conductivity and lattice matching with diverse functional perovskite oxides, SRO thin films have been widely used as model systems to investigate correlated-electron transport and as conducting electrodes in oxide heterostructures.[3] In particular, epitaxial SRO films exhibit a wide variety of phenomena, including strain-tunable ferromagnetism,[6,7] metal-insulator transitions in ultrathin layers,[8–10] and a large anomalous Hall effect governed by the Berry curvature, which is highly sensitive to band structure and spin-orbit coupling.[11,12] Recently, the observation of quantum transport phenomena of Weyl fermions, including a high quantum mobility of about 10,000 cm$^2$V$^{-1}$s$^{-1}$ and linear positive magnetoresistance (MR), in SRO has also attracted considerable interest as a magnetic Weyl semimetal,[13,14] which serves as a key platform bridging oxide electronics and topological transport.

Research on SRO has primarily focused on (001)-oriented epitaxial films on a SrTiO$_3$ (STO) (001) substrate, in which atomically flat surfaces and low defect densities are routinely achieved, thereby enabling high-quality films. By contrast, in (111)-oriented perovskite oxides, the surface structural symmetry changes from cubic to triangular, and the (111) surface can display a buckled honeycomb-like arrangement of surface atoms when viewed along [111]. This geometry provides a fertile ground for many quantum phenomena, including tunable band-structure topology,[15] and the quantum anomalous Hall (QAH) state.[16,17] For SRO specifically, theoretical work on (111)-oriented SRO-based heterostructures predicts the QAH Hall state in ultrathin ferromagnetic bilayers,[18] which requires atomically smooth interfaces and highly ordered films to realize. There have also been reports of a strain-induced high-spin state (4 $\mu_B$ per Ru$^{4+}$) on the (111) plane, prompting a long-standing debate over its existence.[19–24] Motivated by such prospects, considerable effort has been devoted to improving surface preparation of an STO (111) substrate, establishing single-terminated, atomically flat terraces by optimized chemical etching and annealing,[25–28] and to developing growth protocols that enable layer-by-layer or near-layer-by-layer deposition of SRO (111) on the STO (111) substrate.[29] Nevertheless, the (111) plane is thermodynamically unstable due to its high polarity and large bond density per unit area, which has long hampered the epitaxy of high-quality (111)-oriented SRO.[30] Consistent with this, the maximum residual resistivity ratio (RRR) reported for SRO (111) films on STO (111) is only ~9,[21] far below the >80 achieved on SRO (001) films on STO (001).[3,31] Therefore, there is a strong need to fabricate high-quality SRO (111) films and to systematically elucidate the correlations between their structure, magnetism, and transport properties for utilizing topological property in oxide electronics.

In this study, we grew SRO (111) thin films with thicknesses $t$ = 1.2–60 nm on STO (111) substrates by machine-learning-assisted molecular beam epitaxy (ML-MBE) and characterized their structure, magnetism, and magnetotransport properties. High-resolution X-ray diffraction reciprocal space mapping (HRXRD-RSM) and cross-sectional scanning transmission microscopy (STEM) confirmed epitaxy with in-plane coherent strain for $t$ = 10 and 20 nm, and strain relaxation at $t$ = 60 nm. The 60-nm thick film reached a RRR of 45.5. The temperature ($T$)-dependent longitudinal resistivity $\rho_{xx}$ exhibited Fermi-liquid behavior ($\rho_{xx} \propto T^2$) below ~15 K for $t$ = 10, 20, and 60 nm, indicating that these films are suitable for probing intrinsic transport; accordingly, we performed detailed magnetotransport measurements on them. The robust, non-saturating linear positive MR was observed up to 14 T and the anomalous Hall effect (AHE) was analyzed by scaling to separate intrinsic (Karplus–Luttinger) and extrinsic (side-jump) contributions. Comparison between coherently strained ($t$ = 10, 20 nm) and relaxed ($t$ = 60 nm) films shows that the (111) epitaxial strain systematically tunes the relative weight of the AHE mechanisms. Superconducting quantum interference device (SQUID) magnetometry together with soft X-ray absorption spectroscopy (XAS)/X-ray magnetic circular dichroism (XMCD) at Ru $M_{2,3}$ and O $K$-edges including the sum-rule analysis indicated an intrinsically low-spin Ru ground state for both coherently strained and relaxed films.

## II. METHODS

We grew epitaxial SRO films with $t$ of 1.2, 2, 10, 20, and 60 nm on STO (111) substrates in an ML-MBE system[32–34] equipped with multiple e-beam evaporators for Sr and Ru. To obtain atomically smooth surfaces, the STO (111) substrates were etched in an HCl–HNO₃ solution prepared by mixing HCl (36%) and HNO₃ (61%) in a 3:1 volume ratio,[26,28] then annealed at 1000°C for 10 h in an oxygen atmosphere. The oxidation during growth was carried out with a mixture of ozone ($O_3$) and $O_2$ gas (~15% $O_3$ + ~85% $O_2$). The growth rate of 1.1 Å/s was deduced from the STEM-measured thickness of a thick film ($t$ = 60 nm) and the deposition time. All the SRO films were prepared under the same growth conditions: at a substrate temperature of 818°C, a Sr flux rate of 0.98 Ås$^{-1}$, and a Ru flux rate of 3.5 Ås$^{-1}$. The growth parameters were optimized by Bayesian optimization,[32,35] a machine learning technique, with which we achieved a high RRR of 45.5 for the SRO film with $t$ = 60 nm as described in Sec. III.C. Detailed information on the machine learning model and the optimization procedure is described in Ref. 32. Further information about the MBE setup is described elsewhere.[31,36]

Magnetization measurements were performed by a SQUID (MPMS-3, Quantum Design). For Hall measurements, we deposited Ag electrodes on an SRO surface. Then, we fabricated a 180×300 μm$^2$ Hall bar by photolithography and Ar ion milling. Magnetotransport measurements were performed by a DynaCool physical property measurement system (PPMS, Quantum Design). X-ray diffraction (XRD) measurements were performed using a Bruker D8 Discover. The X-ray wavelength was 1.54060 Å (Cu K$\alpha_1$ line source) with a germanium (220) monochromator. Atomic force microscopy (AFM) measurements were performed using a Bruker Dimension FastScan AFM. High-angle annular dark-field (HAADF) and annular bright-field (ABF) STEM measurements were performed using a JEOL JEM-ARM 200F microscope. All the crystallographic indices mentioned in this work are defined in cubic (or pseudo-cubic) notation.

The XAS and XMCD measurements were performed at the helical undulator beamline BL25SU of SPring-8.[37,38] The monochromator resolution $E/\Delta E$ was over 5,000 at the Ru $M_{2,3}$ edges. The spot size of the designed beam was 10 × 100–200 μm$^2$. For the XMCD measurements, absorption spectra for circularly polarized X-rays with the photon helicity parallel ($\mu^+$) or antiparallel ($\mu^-$) to the spin polarization were obtained by reversing the photon helicity at each photon energy $h\nu$ and recording them in the total-electron-yield (TEY) mode. The measurement temperature was 18 K. All the XAS and XMCD measurements were performed at 1.92 T, the maximum applied magnetic field in the system. The $\mu^+$ and $\mu^-$ spectra at the Ru $M_{2,3}$ edges and O $K$ edge were taken under both positive and negative applied magnetic fields and averaged to eliminate spurious dichroic signals. To estimate the integrated values of the XAS spectra at the Ru $M_{2,3}$ edges, hyperbolic tangent functions as background were subtracted from the spectra. The external magnetic fields were applied perpendicular to the sample surface.

**III. RESULTS AND DISCUSSION**

**A. Crystallographic analyses**

Figure 1 shows the HRXRD-RSM scans around the STO (114) reflection to reveal the in-plane and out-of-plane lattice constants and strain state of the SRO (111) films. At $t = 20$ nm, the position of the film peak along the horizontal $q_{\bar{5}2\bar{7}}$ ($q_x$) axis is the same as that of the substrate peak, indicating that the SRO (111) film is coherently strained under in-plane compressive strain on the STO (111) substrate. The SRO film with $t = 20$ nm has the same in-plane lattice constant of SrTiO$_3$ ($d_{110} = a_{STO}/\sqrt{2} = 2.762$ Å),[26] accompanied by an out-of-plane spacing of 2.281 Å. Using the Poisson's ratio of SRO ($\nu = 0.33$), the in-plane compressive strain relative to bulk SRO (-0.64%) predicts an out-of-plane expansion of 0.63%, yielding an out-of-plane spacing of 2.283 Å. This agrees with the measured 2.281 Å, confirming coherent strain in the SRO (111) film. In contrast, at $t = 60$ nm, the SRO peak splits into two; the film peaks are displaced from the substrate peak along $q_x$, indicating lattice relaxation and the formation of two distinct domains, possibly arising from relaxation-induced crystallographic tilts.[39]

We further confirmed the atomically flat surface and single-crystalline epitaxial growth of the coherently strained SRO film with $t = 20$ nm by AFM and STEM measurements. The AFM image of the film surface (Fig. 2) shows a surface morphology composed of flat terraces and molecular steps with a height of ~0.45 nm, corresponding to twice the (111) interplanar spacing ($2d_{111}$) in the pseudocubic lattice of SRO. The surface exhibits a low root-mean-square roughness of 0.19 nm, consistent with a two-dimensional layer-by-layer growth mode. Figure 3 shows atomic-resolution HAADF- and ABF-STEM images of the SRO film with $t = 20$ nm obtained with the electron beams incident along the $[11\bar{2}]$ direction. Since the intensity in the HAADF-STEM image is proportional to ~$Z^n$ (n ~ 1.7–2.0, and Z is the atomic number),[40] the brighter spheres and darker ones in Fig. 3(c) are assigned to Ru- (Z = 44) and Sr- (Z = 38) occupied columns, respectively. The ABF-STEM images [Figs. 3(b) and 3(d)] represent the atomic arrangement of oxygen since the oxygen is emphasized in annular bright-field ABF-STEM images.[41] At a glance, it is evident that a single-crystalline SRO film with an abrupt substrate/film interface was grown epitaxially on a (111) STO substrate. The film is coherently strained to the substrate (no detectable lattice relaxation), consistent with the HRXRD-RSM results [Fig. 1(b)]. In the inset of Fig. 3(c), the Sr zigzag pattern characteristic of the orthorhombic *Pbnm* structure with $a^-a^-c^+$ rotation in Glazer notation[42] is observed,[43,44] indicating that the SRO film with $t = 20$ nm has a coherently compressively strained orthorhombic *Pbnm* structure.

B. Magnetic properties

Figure 4(a) shows the temperature dependence of the magnetization for the SRO (111) films with $t$ = 20 nm and 60 nm. A magnetic field of 500 Oe was applied perpendicular to the film plane. The Curie temperatures ($T_C$) are 160 K ($t$ = 20 nm) and 161 K ($t$ = 60 nm), which are comparable to that of bulk SRO (160–165 K).[45] This contrasts with SRO (001) films on STO (001), where compressive epitaxial strain reduces $T_C$ to 152 K;[7,46–48] here, despite the SRO (111) film with $t$ = 20 nm being coherently strained in-plane, its $T_C$ remains essentially unchanged. Figure 4(b) shows magnetization hysteresis loops of the SRO (111) films with $t$ = 20 and 60 nm, measured at 2.5 and 4 K, respectively, with the magnetic field applied perpendicular to the film plane. Rectangular hysteresis loops with saturation magnetization values of 1.18 $\mu_B$ per pseudocubic unit cell ($t$ = 20 nm) and 1.38 $\mu_B$ per pseudocubic unit cell ($t$ = 60 nm) were observed. The values of the saturation magnetization are typical for bulk and SRO films on STO (001).[3,49] A magnetic moment exceeding 3 $\mu_B$ per pseudocubic unit cell, corresponding to the high-spin configuration $t_{2g}(3\uparrow)e_g(1\uparrow)$, has often been reported for SRO (111) films grown on STO (111).[19,20,23,24] Grutter et al. further showed that this putative high-spin state disappears as lattice relaxation proceeds with increasing thickness, attributing it to compressive epitaxial strain from the STO (111) substrate.[19] However, in our high-quality SRO (111) films, neither the relaxed thick film ($t$ = 60 nm) nor the coherently in-plane compressed thin film ($t$ = 20 nm) exhibits an enhanced magnetic moment indicative of a high-spin state. Notably, in previous studies in which a high-spin state was reported, all the SRO (111) films showed low RRR below 8, indicating that the films have substantial defects. These results suggest that the large moments (> 3 $\mu_B$ per pseudocubic unit cell) are unlikely to be intrinsic to SRO (111), but are more plausibly associated with defects and/or microstructural inhomogeneity in conjunction with (111)-induced strain.

To obtain the detailed insight of the electronic structure and magnetic properties of the SRO (111) films, we employed XAS and XMCD, both of which provide element-specific sensitivity.[50–53] For SRO, numerous studies have established that applying the XMCD sum rules to the Ru $M_{2,3}$-edge XAS/XMCD spectra yields the spin ($m_{spin}$) and orbital ($m_{orb}$) magnetic moments of Ru.[20,54–57] Moreover, an orbital magnetic moment on oxygen, arising from strong Ru 4$d$-O 2$p$ hybridization,[55,58] has been detected, making XAS/XMCD a standard approach for evaluating the Ru 4$d$ and O 2$p$ electronic structures and magnetic properties in an element-specific way. Figure 5(a) shows XAS and XMCD spectra at the Ru $M_{2,3}$ edge for the SRO (111) film with $t$ = 60 nm, measured at 18 K under a magnetic field $\mu_0H$ of 1.92 T applied perpendicular to the film plane. The Ru $M_3$ and $M_2$ XAS peaks, arising from transitions from the Ru $3p_{3/2}$ and $3p_{1/2}$ core levels to the Ru $4d$ states, appear at around 463.5 and 485 eV, respectively, with clear corresponding XMCD signals indicating a ferromagnetic moment on Ru. As shown in Fig. 5(b), the O $K$-edge XAS and XMCD probe the unoccupied electronic structures hybridized with the O $2p$ states and the O $2p$ orbital magnetic moments. The absorption peak at 529 eV arises from the transition to the coherent part of the Ru $4d$ $t_{2g}$ states, while features at 529.5–534 eV correspond to the incoherent part of the Ru $4d$ $t_{2g}$ states and $e_g$ states, as reported by Refs. [54,55]. The O $K$-edge XMCD at the Ru $4d$ $t_{2g}$ peak evidences a sizable O $2p$ orbital moment, arising from strong Ru $4d$-O $2p$ hybridization and charge transfer.

Figures 5(c) and 5(d) show Ru $M_{2,3}$-edge and O $K$-edge XMCD spectra normalized at 462.2 eV and 592.2 eV, respectively, for the SRO films with $t$ = 10, 20, and 60 nm. In the normalized Ru $M_{2,3}$-edge XMCD spectra, the smaller $M_2/M_3$ intensity ratio in the coherently strained thin films ($t$ = 10 and 20 nm) than the relaxed thick film ($t$ = 60 nm) indicates that the ratio of orbital to spin magnetic moments, $m_{orb}/m_{spin}$, is modulated by the (111) in-plane epitaxial strain, as will be discussed in detail using the XMCD sum-rule analysis. By contrast, the normalized O $K$-edge XMCD spectra are essentially identical across all films, indicating negligible changes in the oxygen electronic and magnetic states.

We determined the orbital magnetic moment $m_{orb}$ and the spin magnetic moment $m_{spin}$ of the $Ru^{4+}$ $4d$ states using the XMCD sum rules as follows:[50–52]

$$m_{orb} = -\frac{4(10-n_{4d})}{3r} \int_{M_2+M_3} (\mu^+ - \mu^-)dE, \qquad (1)$$

$$m_{spin} + 7m_T = -\frac{2(10-n_{4d})}{r}[\int_{M_3}(\mu^+ - \mu^-)dE - 2\int_{M_2}(\mu^+ - \mu^-)dE]. \qquad (2)$$

Here, $r = \int_{M_2+M_3}(\mu^+ + \mu^-)dE$, and $n_{4d}$ is the number of electrons in $4d$ orbitals, which is assumed to be four. For ions in octahedral symmetry, the magnetic dipole moment $m_T$ is a small number compared to $m_{spin}$.[59] By dividing equation (1) by equation (2), $m_{orb}/m_{spin}$ is expressed by

$$\frac{m_{orb}}{m_{spin}} = \frac{2}{3} \frac{\int_{M_2+M_3}(\mu^+-\mu^-)dE}{\int_{M_3}(\mu^+-\mu^-)dE - 2\int_{M_2}(\mu^+-\mu^-)dE}. \qquad (3)$$

Thus, we can obtain $m_{orb}/m_{spin}$ directly from the XMCD spectra without an assumption of the $n_{4d}$ value. Table 1 summarizes $m_{spin}$, $m_{orb}$, and $m_{orb}/m_{spin}$ for the SRO (111) films with $t$ = 10, 20, and 60 nm obtained from the Ru $M_{2,3}$-edge XAS and XMCD spectra at 18 K under a magnetic field $\mu_0 H$ of 1.92 T applied perpendicular to the film plane. The total magnetic moment, $m_{total}$ =

$m_{spin} + m_{orb}$ is 0.59 $\mu_B$/Ru for $t = 60$ nm and 0.57 $\mu_B$/Ru for $t = 20$ nm, which are smaller than the magnetization measured by SQUID (1.18 and 1.38 $\mu_B$ per pseudocubic unit cell, respectively). This discrepancy likely originates from the magnetization of the O 2p electrons, as verified by the O K-edge XMCD spectra. Assuming that oxygen contributes 30% of the total magnetization,[60] as reported for bulk SRO, the sum of the Ru and O ion's magnetizations estimated from the $m_{total}$ are 0.84 ($t = 60$ nm) and 0.81 ($t = 20$ nm) $\mu_B$ per pseudocubic unit cell, which are closer to the SQUID values. For the relaxed thick film ($t = 60$ nm), $m_{orb}$ is negligibly small, consistent with the orbital-moment quenching reported for bulk SRO.[55] In contrast, for the coherently in-plane compressed thin films ($t = 10$ and 20 nm), a slight but finite orbital moment emerges with $m_{orb}/m_{spin} = 0.013$, suggesting that the (111) in-plane coherent strain modifies the electronic structure.

**C. Magnetotransport**

Figure 6(a) shows temperature-dependent longitudinal resistivity $\rho_{xx}$ of SRO (111) films with $t = 1.2$–60 nm. When $t = 10$–60 nm, the $\rho_{xx}$ decreases with decreasing temperature, indicating that these films are metallic over the entire temperature range. On the other hand, for $t = 2$ nm, the $\rho_{xx}$ decreases as the temperature decreases from room temperature but starts to increase below 69 K, indicating insulating behavior plausibly due to weak localization in the low-temperature region.[3] For $t = 1.2$ nm, an insulating, monotonic increase in resistivity is observed over the entire temperature range. In previous work on SRO (111) films on STO (111) substrates grown by pulsed laser deposition, the 2.3-nm-thick SRO became insulating,[30] while our high-quality SRO (111) film with $t = 2$ nm remains conductive. This indicates that fine-tuning the growth conditions by ML-MBE suppresses disorders due to Ru vacancies and other defects.[61] The $\rho_{xx}$ value of 240 $\mu\Omega$cm at 300 K for the thick film ($t = 60$ nm) is higher than that of thick SRO (001) films on STO (001) (130-150 $\mu\Omega$cm),[3] suggesting that SRO (001) films are more suitable as a metallic oxide electrode, which can lead to lower Joule heating and energy loss in heteroepitaxially grown devices. Figure 6(b) shows the thickness $t$ dependence of the RRR values, defined as $\rho_{xx}(300\ K)/\rho_{xx}(2\ K)$. The RRR of 45.5 at $t = 60$ nm is the highest value for SRO (111) films, suitable for revealing the intrinsic electronic structure and physical properties. The RRR decreases with decreasing $t$, suggesting that there is disorder near the interface between SRO and STO substrates, which is rather insensitive to cross-sectional STEM measurements.

Figure 6(c) shows the first derivative of the longitudinal resistivity, $d\rho_{xx}/dT$, which we use to determine $T_C$ of the films. We define $T_C$ as the temperature at which the $d\rho_{xx}/dT$ curve exhibits a peak. $T_C$ decreases with decreasing thickness $t$, likely due to disorders near the interface that degrade the exchange interactions in SRO [Fig. 6(d)]. The inset of Fig. 6(e) shows the $\rho_{xx}$ vs. $T^2$ plot with a linear fit (black line) for $t = 60$ nm as a representative example. The SRO films with $t = 10$–$60$ nm follow a $T^2$ dependence ($\rho_{xx} \propto T^2$) at low temperatures, as expected for a Fermi liquid in which electron-electron scattering dominates transport. From this, we identify a temperature below which the transport is intrinsic, which we denote $T_F$. Figure 6(e) shows $T_F$ as a function of $t$. Both the coherently strained SRO (111) films with $t = 10$ and $20$ nm and the strain-relaxed film with $t = 60$ nm exhibit Fermi-liquid transport at low temperatures, indicating that intrinsic magnetotransport properties can be probed in this regime.

Figure 7(a) shows MR [$(\rho_{xx}(\mu_0H)-\rho_{xx}(0\ T))/\rho_{xx}(0\ T)$] for the SRO (111) films with $t = 10$, $20$, and $60$ nm at 2 K with a magnetic field applied perpendicular to the film plane. In all the films, anisotropic magnetoresistance (AMR), which is proportional to the relative angle between the electric current and the magnetization, is clearly observed.[62] The peak positions of the AMR for the SRO (111) films with $t = 20$ and $60$ nm correspond to the coercive fields in Fig. 4(b). Importantly, all the films show a linear positive MR with no signature of saturation even up to 14 T, as commonly observed in Weyl semimetals.[13,63–65] For SRO (001) films on STO (001), linear positive MR, a hallmark of Weyl-fermion quantum transport, is observed only for RRR ≳ 20.[66] In the SRO (111) films, however, the same signature remains visible even at RRR = 7.8, suggesting that topologically nontrivial Weyl-fermion transport in SRO(111) is more robust against defects.

Figure 7(b) shows the magnetic field $\mu_0H$ dependence of the Hall resistivity $\rho_{xy}(\mu_0H)$ for the SRO (111) films with $t = 10$, $20$, and $60$ nm at 2 K. The $\rho_{xy}(\mu_0H)$ curves for the SRO films with $t = 60$ nm at 2 K exhibit a negative slope at low fields and a positive slope at high fields, indicating a pronounced nonlinearity. Such behavior is reminiscent of Weyl semimetals with multiple carrier types[63] and has also been reported for high-quality SRO (001) films.[13] For $t = 10$ nm, the magnetization hysteresis curve characteristic of the AHE is clearly observed, and it dominates Hall resistivity at near-zero magnetic fields $\rho_{xy}(0\ T)$. In contrast, for $t = 20$ and $60$ nm, the AHE components become smaller due to the small $\rho_{xx}$ values, following the well-established scaling of $\rho_{xy}(0\ T)$ proportional to $\rho_{xx}^2$ at low enough temperatures.[67,68]

The origins of the anomalous Hall effect include intrinsic origins such as the Karplus-

Luttinger mechanism,[69] and extrinsic origins such as the side-jump[70] and skew-scattering mechanisms.[71,72] To investigate the origin of the AHE in SRO (111) thin films, we measured the temperature dependence of the AHE. Figure 8(a) plots the normalized remanent Hall resistivity $\rho'_{xy}(0\text{ T})$ as a function of the normalized longitudinal resistivity $\rho'_{xx}$, pairing the two values at each temperature, for the SRO (111) films with $t$ = 10, 20, and 60 nm. Here, for the scaling analysis, we define $\rho'_{xy}(0\text{ T})$ as the remanent Hall resistivity normalized by its maximum absolute value, and $\rho'_{xx}$ as the longitudinal resistivity normalized by its value at the temperature when $\rho_{xy}(0\text{ T})$ changes sign. This sign change reflects the cancellation of competing AHE contributions (see the model analysis below), and thus provides a practical reference point for comparing different samples, following the well-established AHE analytical procedure.[67] The $\rho'_{xy}(0\text{ T})$ curves show positive peaks around $T_\text{C}$, then change from positive to negative with decreasing $\rho'_{xx}$. At even lower temperatures, $\rho'_{xy}(0\text{ T})$ values increase again. This trend in SRO is commonly attributed to a combination of the intrinsic Karplus-Luttinger and extrinsic side-jump mechanisms.[67]

Notably, for $\rho'_{xx}$ < 0.6, the coherently strained films ($t$ = 10 and 20 nm) and the strain-relaxed film ($t$ = 60 nm) exhibit distinct scaling of $\rho'_{xy}(0\text{ T})$ with $\rho'_{xx}$. At $\rho'_{xx}$ = 0.384, the separation in $\rho'_{xy}(0\text{ T})$ is maximal, reaching 0.104. This indicates different contributions to the AHE from the Karplus-Luttinger and the side-jump mechanisms. To evaluate these contributions quantitatively, we fit $\rho'_{xy}(0\text{ T})$ vs $\rho'_{xx}$ curves by scaling law expressed as follows:[67]

$$\rho'_{xy}(0\text{ T}) = \frac{a_1}{\Delta^2 + a_2(\rho'_{xx})^2}(\rho'_{xx})^2 + a_3(\rho'_{xx})^2, \qquad (4)$$

where $a_1$, $a_2$, $a_3$ and $\Delta$ are fitting parameters, which are related to the band structure of SRO. The first term describes the contribution from the Karplus-Luttinger mechanism, and the second term describes that from the side-jump scattering. Figures 8(b) and 8(c) show $\rho'_{xy}(0\text{ T})$ vs. $\rho'_{xx}$ curves for the strain-relaxed film ($t$ = 60 nm) and the coherently strained films ($t$ = 10 and 20 nm), respectively. In both cases, the data are well reproduced by the scaling relation in Eq. (4), indicating that the AHE in SRO (111) films has a common origin arising from a combination of the Karplus-Luttinger and side-jump mechanisms. Different sets of fitting parameters are obtained for the strain-relaxed and coherently strained films, which enable us to evaluate the respective contributions of the Karplus-Luttinger mechanism and side-jump scattering in each case. The parameters for the strain-relaxed film (the coherently strained films) obtained by the fitting are $a_1$ = −5.44 (−19.44), $a_2$ = 0.83 (1.64), $a_3$ = 4.61 (6.27), and $\Delta$ = 0.67 (1.27). Figure 8(d) shows the $\rho'_{xx}$ dependence of the Karplus-Luttinger-to-side-jump coefficient ratio from Eq. (4) for the strain-relaxed film and the coherently strained films. In both cases, the contribution from the Karplus-Luttinger term becomes more dominant with decreasing temperature. Our analysis indicates that (111) epitaxial strain serves as a practical knob to tune the AHE mechanism balance in SRO (111): in coherently strained films, the intrinsic Karplus-Luttinger term is relatively suppressed compared with the strain-relaxed film, shifting the balance toward the extrinsic side-jump channel. While our scaling analysis captures strain-dependent trends in the relative weights of intrinsic- and extrinsic-like channels, microscopic attribution remains open. Possible origins include strain-modified band topology/Berry curvature and disorder-related scattering pathways; however, discriminating among them requires future microstructural/spectroscopic studies beyond the scope of this work.

**IV. CONCLUSION**

We grew high-quality SRO (111) films on STO (111), achieving a high RRR of 45.5 at $t$ = 60 nm and exhibiting Fermi-liquid transport below ~15 K across coherently strained ($t$ = 10 and 20 nm) and relaxed ($t$ = 60 nm) films, enabling reliable analyses of intrinsic magnetotransport. Element-specific XAS/XMCD together with SQUID magnetometry demonstrates an intrinsically low-spin Ru ground state for both strain states, suggesting that previously reported high-spin behavior is not intrinsic to SRO (111) but is likely extrinsic, arising from defects together with (111)-induced strain. Magnetotransport reveals robust, non-saturating linear positive MR up to 14 T and an AHE with temperature scaling that separates intrinsic (Karplus–Luttinger) and extrinsic (side-jump) contributions; notably, (111) epitaxial strain systematically tunes their relative weights. Together, these results position SRO (111) as a materials platform for (111)-oriented oxide heterostructures and spin-topological devices.

**ACKNOWLEDGMENTS**

This work was partially supported by the Spintronics Research Network of Japan (Spin-RNJ). The synchrotron radiation experiments were performed at the BL25SU of SPring-8 with the approval of the Japan Synchrotron Radiation Research Institute (JASRI) (Proposal No. 2025A1361).

**AUTHOR DECLARATIONS**
**Conflict of interest**

The authors have no conflicts to disclose.

**Author Contributions**

H.S. and Y.K.W. contributed equally to this work.

**H. Shiratani:** Conceptualization (supporting); Validation (equal); Investigation (equal); Writing – Original Draft (equal); Writing – Review & editing (equal). **Y. K. Wakabayashi:** Conceptualization (lead); Validation (equal); Investigation (equal); Supervision (lead); Writing – Original Draft (equal); Writing – Review & editing (equal). **Y. Krockenberger:** Investigation (supporting); Writing – Review & editing (supporting). **M. Kobayashi:** Investigation (supporting); Writing – Review & editing (supporting). **K. Yamagami:** Investigation (supporting); Writing – Review & editing (supporting). **T. Takeda:** Investigation (supporting); Writing – Review & editing (supporting). **H. Yamamoto:** Writing – Review & editing (supporting). **S. Ohya:** Writing – Review & editing (supporting). **M. Tanaka:** Writing – Review & editing (supporting). **Y. Taniyasu:** Writing – Review & editing (supporting).

**DATA AVAILABILITY**

The data that support the findings of this study are available from the corresponding author upon reasonable request.

**Figures and figure captions**

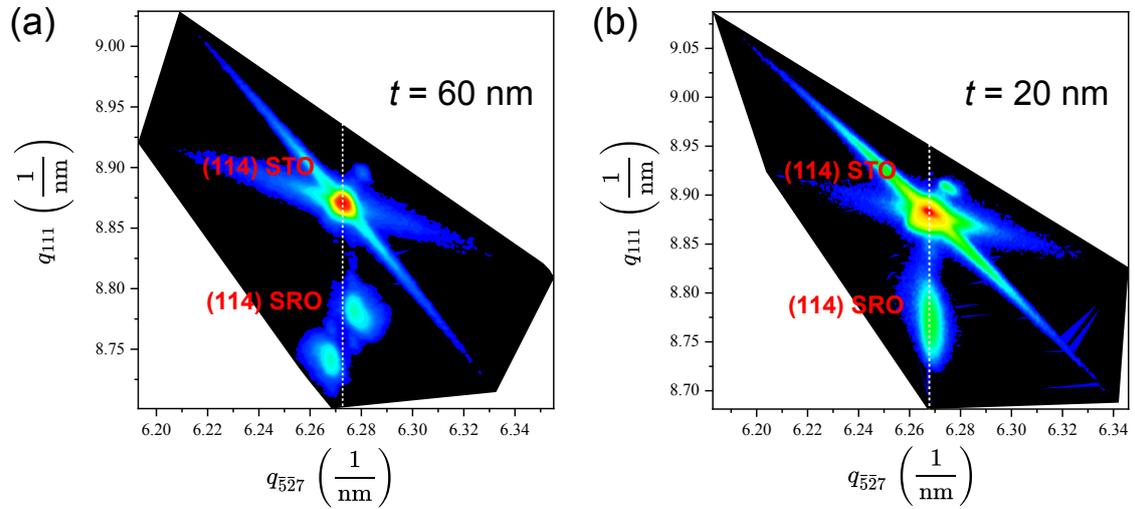

**Fig. 1.** HRXRD-RSM scans near the STO (114) reflection for the SRO (111) films with $t$ = (a) 60 nm and (b) 20 nm. White dashed lines indicate the $q_{\bar{5}\bar{2}7}$ ($q_x$) position of the STO (114) reflections.

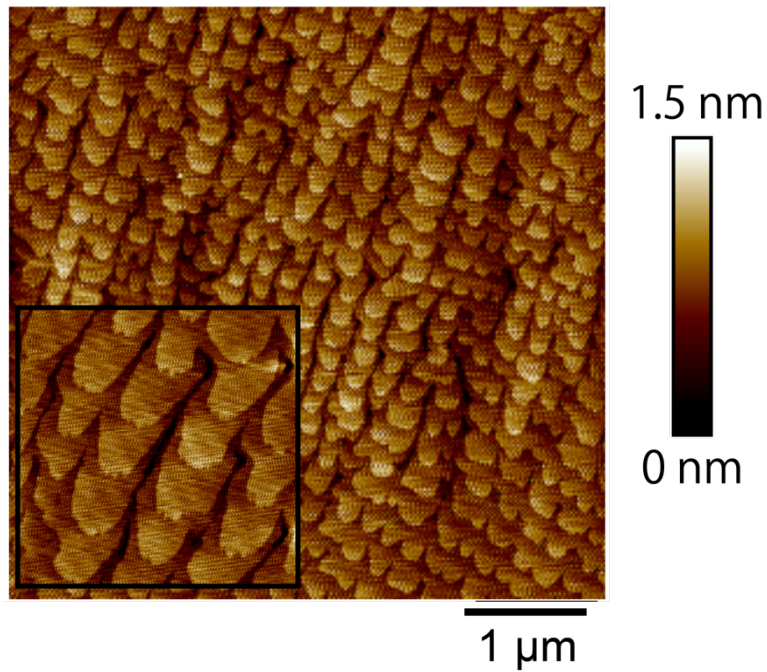

**Fig. 2.** AFM image of the SRO (111) film with $t$ = 20 nm. The inset shows a magnified image with the scan area of $1 \times 1$ $\mu m^2$, highlighting the step and terrace structure.

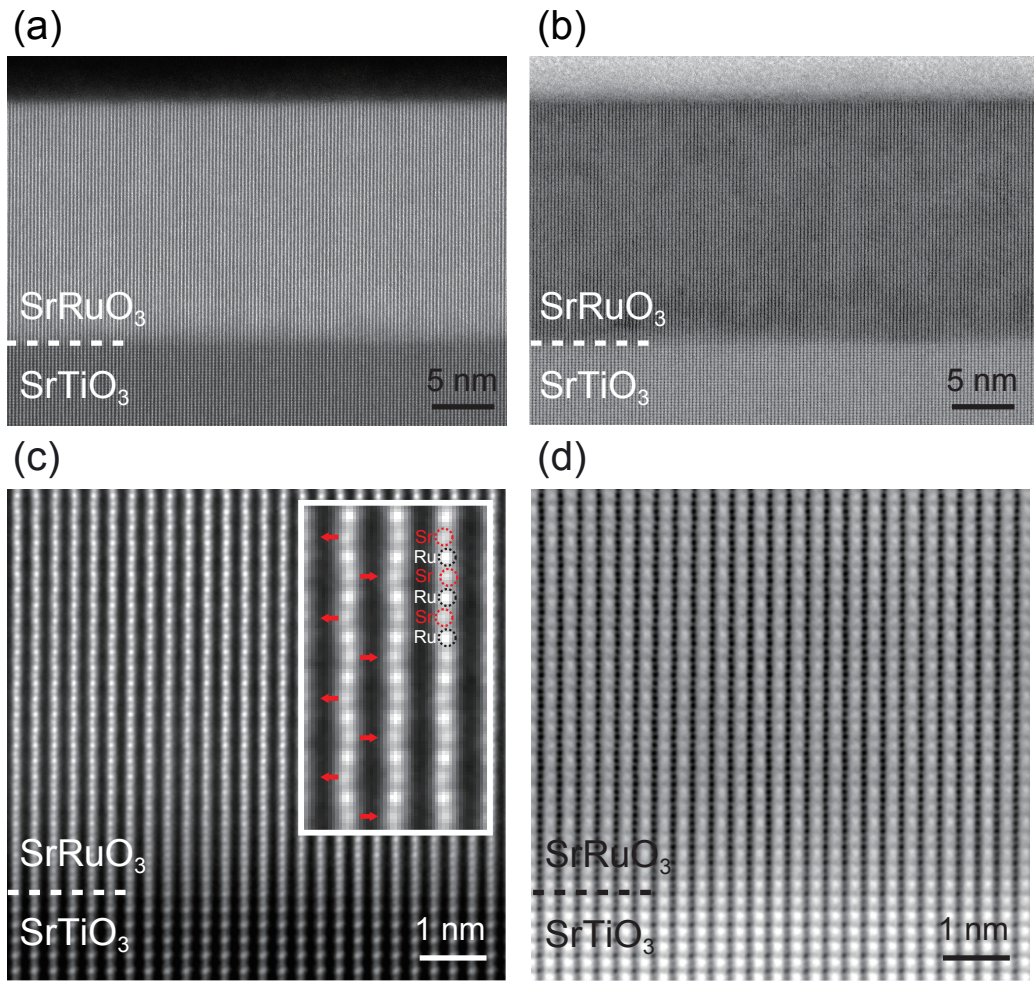

**Fig. 3.** (a),(c) HAADF-STEM and (b),(d) ABF-STEM images of the SRO (111) film with $t = 20$ nm along the $[11\bar{2}]$ direction. (c) and (d) show magnified, high-resolution images of the interfacial regions in (a) and (b), respectively. Dashed lines indicate the interface between the SRO film and the STO substrate. In (c), insets show magnified views highlighting the Sr zigzag pattern, with arrows marking the displacements of Sr atoms.

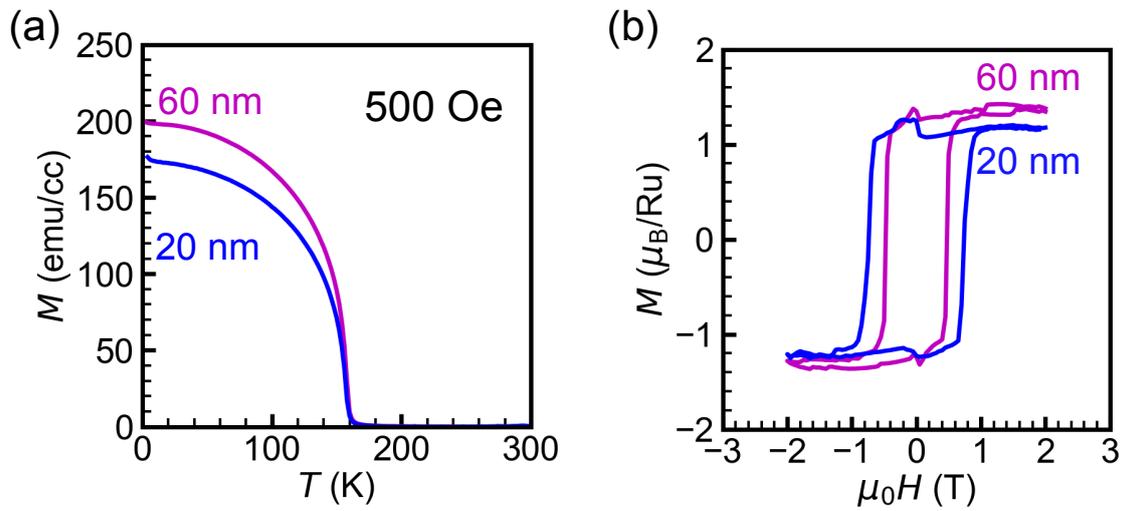

**Fig. 4.** (a) Temperature $T$ dependence of the magnetization $M$ for the SRO (111) films with $t$ = 20 and 60 nm. A magnetic field of 500 Oe was applied perpendicular to the film plane. (b) $M$-$\mu_0H$ curves for the SRO (111) with $t$ = 20 nm at 2.5 K and $t$ = 60 nm at 4 K. A magnetic field was applied perpendicular to the film plane.

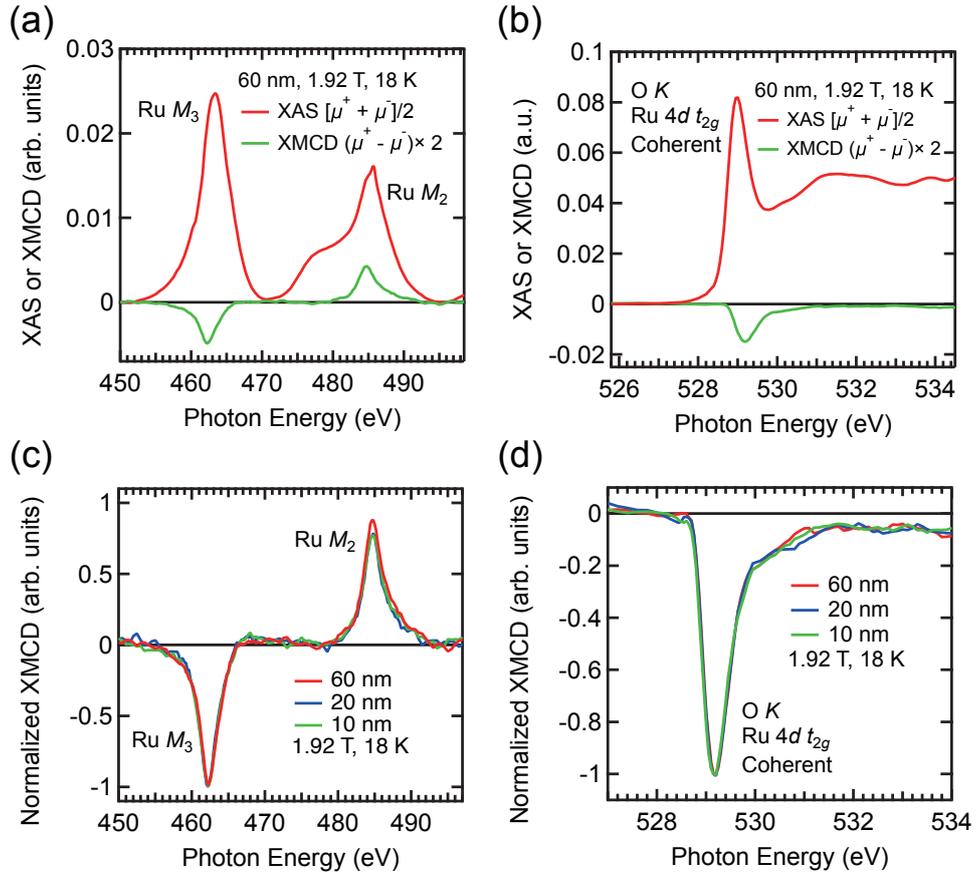

**Fig. 5.** (a) Ru $M_{2,3}$-edge XAS and XMCD spectra and (b) O $K$-edge XAS and XMCD spectra for the SRO film with $t$ = 60 nm at 18 K under a magnetic field $\mu_0H$ of 1.92 T applied perpendicular to the film plane. (c) Ru $M_{2,3}$-edge and (d) O $K$-edge XMCD spectra normalized at 462.2 eV and 592.2 eV, respectively, for the SRO films with $t$ = 10, 20, and 60 nm.

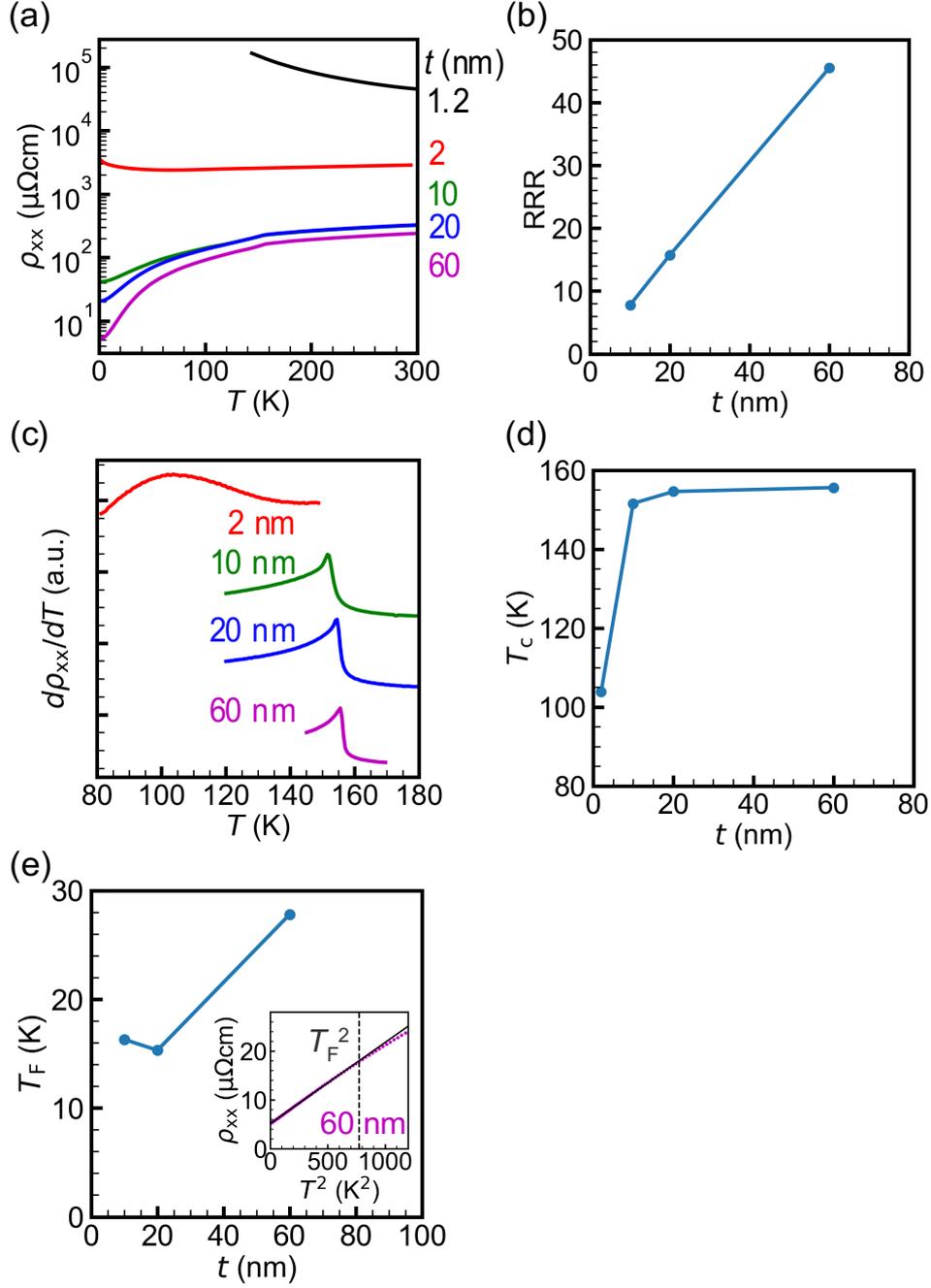

**Fig. 6.** (a) Temperature $T$ dependence of $\rho_{xx}$ for the SRO (111) films with $t$ = 1.2-60 nm. (b) Thickness $t$ dependence of the RRR value. (c) First derivative of the longitudinal resistivity $d\rho_{xx}/dT$ for the SRO (111) films with $t$ = 1.2-60 nm. (d) Thickness $t$ dependence of $T_C$, defined as the peak temperature in (c). (e) Thickness $t$ dependence of $T_F$. The inset shows the $\rho_{xx}$ vs. $T^2$ plot with the linear fit (black line) for the SRO (111) film $t$ = 60 nm. We define $T_F$ as the temperature below which the deviation of $\rho_{xx}$ from the linear fit is within 0.1 μΩ cm.

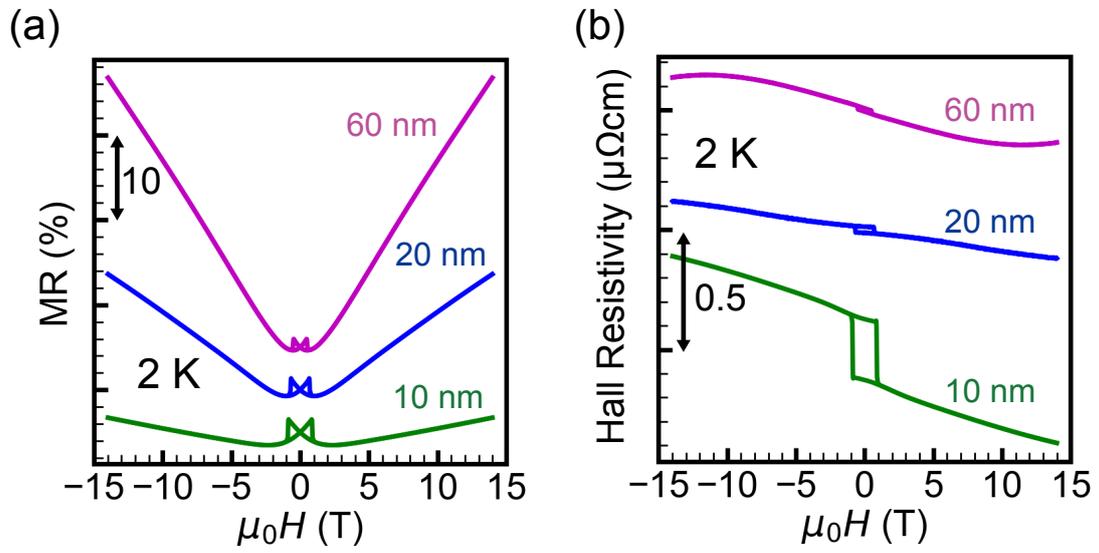

**Fig. 7.** (a) Magnetoresistance MR [($\rho_{xx}(\mu_0H)-\rho_{xx}(0\ T))/\rho_{xx}(0\ T)$] and (b) Hall resistivity $\rho_{xy}(\mu_0H)$ for the SRO (111) films with $t = 10, 20$, and 60 nm at 2 K with a magnetic field applied perpendicular to the film plane.

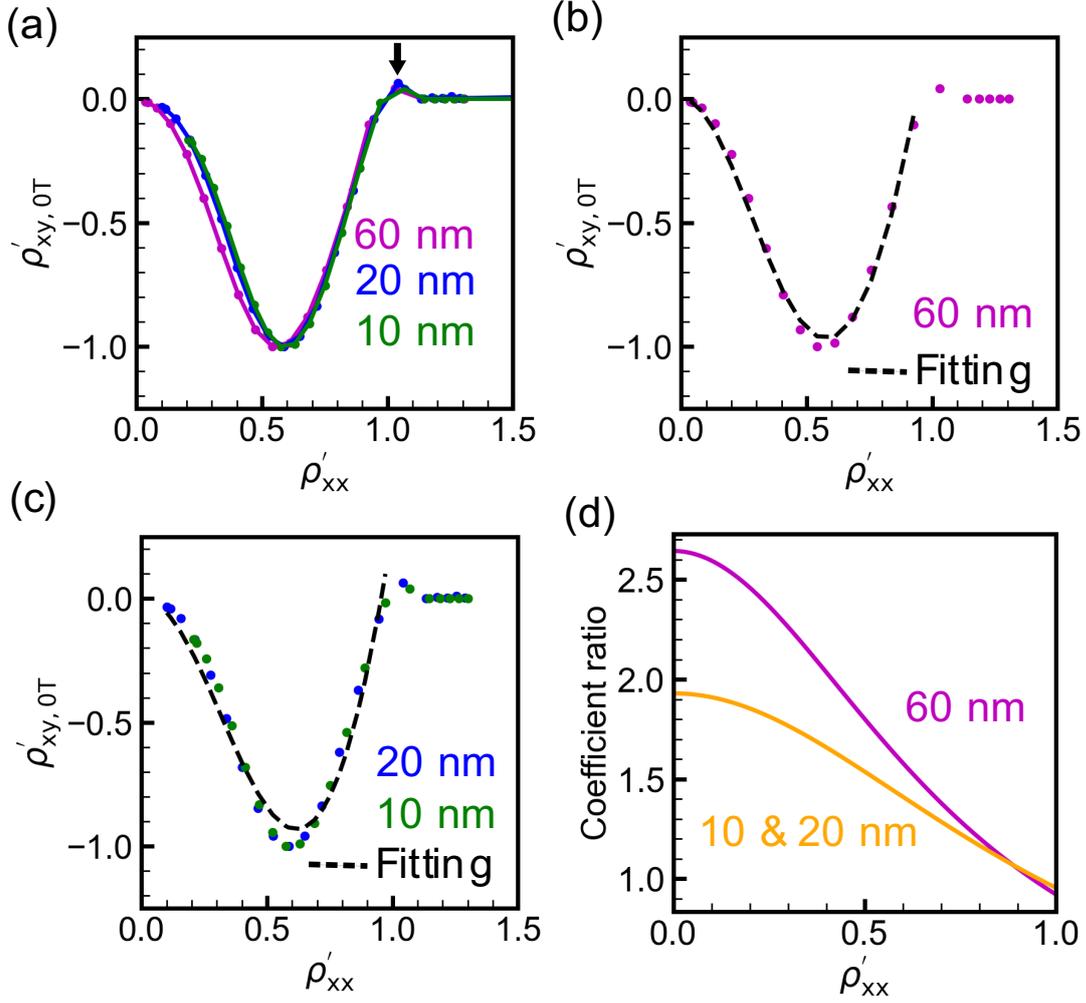

**Fig. 8.** (a) Normalized remanent Hall resistivity $\rho'_{xy,0\mathrm{T}}$ vs. normalized longitudinal resistivity $\rho'_{xx}$ curves for the SRO films with $t$ = 10, 20, and 60 nm. (b),(c) $\rho'_{xy,0\mathrm{T}}$ vs. $\rho'_{xx}$ curves for (b) $t$ = 60 nm and (c) $t$ = 10 nm and 20 nm. In (b) and (c), the dashed lines show scaling fits to Eq. (4). (d) The $\rho'_{xx}$ dependence of the Karplus-Luttinger-to-side-jump coefficient ratio obtained from Eq. (4) for $t$ = 60 nm and $t$ = 10 and 20 nm. Here, the ratio is defined as the absolute value of the coefficient of the Karplus–Luttinger term divided by that of the side-jump term in Eq. (4).

**Table 1**. $m_{spin}$, $m_{orb}$, and $m_{orb}/m_{spin}$ [$\mu_B$/Ru] values estimated from the XMCD sum rules.

| Sample | $m_{spin}$ [$\mu_B$/Ru] | $m_{orb}$ [$\mu_B$/Ru] | $m_{orb}/m_{spin}$ |
|---|---|---|---|
| $t$ = 60 nm | 0.58 ± 0.04 | 0.001 ± 0.002 | 0.002 ± 0.006 |
| $t$ = 20 nm | 0.50 ± 0.04 | 0.007 ± 0.002 | 0.013 ± 0.006 |
| $t$ = 10 nm | 0.60 ± 0.04 | 0.008 ± 0.002 | 0.013 ± 0.006 |